\documentclass[
aps,%
12pt,%
final,%
notitlepage,%
oneside,%
onecolumn,%
nobibnotes,%
nofootinbib,%
superscriptaddress,%
noshowpacs,%
centertags]%
{revtex4}

\begin{document}

\title{Interactions of High Energy Cosmic Rays with \\
Extragalactic Infrared Radiation Background}

\author{\firstname{Edgar}~\surname{Bugaev}}
\email{bugaev@pcbai10.inr.ruhep.ru}
\author{\firstname{Peter}~\surname{Klimai}}
\email{pk@pochta.ru} %
\affiliation{ Institute for Nuclear Research, Russian Academy of
Sciences, 60th October Anniversary Prospect 7a, 117312 Moscow,
Russia }

\begin{abstract}
We consider the modification of extragalactic cosmic ray spectrum
caused by cosmic ray interactions with infrared background photons
which are present in the extragalactic space together with relic
photons. It is assumed that cosmic ray spectrum at superhigh
energies has extragalactic origin and is proton dominated.
\end{abstract}

\maketitle

\section{Introduction}
Presence of  cosmic infrared background radiation (CIB) in
extragalactic space has now been confirmed by experiments, and its
intensity is measured (see, e.g., reviews \cite{franceschini,
hauser}), though the precision of these measurements is still not
so high, especially if to compare with those for relic radiation.
Nevertheless, we know enough to estimate some effects caused by
the presence of the extragalactic infrared background. One of such
effects is neutrino production due to high energy cosmic ray
proton interactions with CIB photons, considered in
\cite{stanev,b1,b2}. Here we will focus on the other point: high
energy cosmic ray spectrum modification due to interactions with
CIB photons.

\section{Extragalactic cosmic ray proton spectrum}
For the calculation of cosmic ray (CR) energy spectrum at high
energies, we assume it to be extragalactic and proton dominated
from the energy $E_0=3\times 10^{17} eV$. The sources of these CRs
are assumed to be distributed isotropically and uniformly
throughout the Universe. In this case it is very convenient to use
cosmological transport equation written as
\begin{equation}
\label{eq1} \frac{\partial n (E,z)}{\partial
z}+\frac{\partial}{\partial E} \left[\beta(E,z)n(E,z)\right] -
\frac{3n(E,z)}{1+z}=g(E,z).
\end{equation}
We work in in the continuous energy loss approximation introduced
in \cite{berezinsky} and neglect possible proton absorbtion. In
(\ref{eq1}), $n(E,z)$ is the number density of CR protons with a
given redshift $z$, the function $\beta(E,z)$ is the change of
proton energy in unit interval of $z$,
$$
%\label{eq2}
\beta(E,z)=\frac{dE}{dz} = \frac{E}{1+z} - b(E,z)
\frac{dt}{dz},
$$
where the first term in r.h.s. is due to adiabatic energy losses
(caused by cosmological expansion), $b(E,z)=-dE/dt$, and is simply
connected with the proton cooling rate $t_p^{-1}(E,z)$. For the
case of proton interactions with photon gas it is given by the
formula
$$
t_{p}^{-1}(E,z)=\frac{1}{E}b(E,z)=\frac{c}{2\gamma_p^{2}} \int
\limits_{\epsilon_{th}}^{\infty} d \epsilon_{r}
\sigma(\epsilon_{r}) f(\epsilon_{r}) \epsilon_{r}\int \limits_{
\epsilon_{th}/{2\gamma_p} }^{\infty} d\epsilon
\frac{n^{photon}(\epsilon,z)}{\epsilon^2},
%\label{eq3}
$$
where $\gamma_p=E/m_p$ is proton Lorenz factor, $\epsilon_{r}$ is
the photon energy in the CR proton rest system,
$\sigma(\epsilon_{r})$ is the photoabsorbtion cross section,
$f(\epsilon_{r})$ is the average relative proton energy loss in
$p\gamma$-collision (in the observer system),  $\epsilon_{th}$ is
the threshold of the photoabsorbtion reaction.

Protons lose their energy interacting with photons via $p\gamma\to
\pi X$ and $p\gamma\to pe^{+}e^{-}$ reactions. Here, we suppose
that $\gamma$ can be relic photon (they have black body spectral
distribution with $T\approx 2.7K$) or background infrared photon.

Thus, in our approximation the total cooling rate should be
written as a sum of two components: cooling due to interactions
with relic and infrared components of the extragalactic radiation
background,
$$
%\label{eq4}
t_p^{-1}(E,z) = t_{p,relic}^{-1}(E,z) + t_{p,infr}^{-1}(E,z).
$$

The function $g(E,z)$ in r.h.s. of the kinetic equation
(\ref{eq1}) describes the combined source of extragalactic cosmic
rays. This source function can be written in the form
\begin{equation}
\label{eq5} g(E,z)=\rho(z)\eta(z)f(E)\frac{dt}{dz}.
\end{equation}
Here, $\rho(z)$ is the number density of local CR sources (e.g.,
AGNs) in the proper (physical) volume, $\rho(z)=\rho_0 (1+z)^3$,
$\eta(z)$ is the activity of each local source (the integrated
number of produced particles per second), $\eta(z)=(1+z)^{m}
\eta_0\theta(z_{max}-z)$. Writing this, we assume that the
cosmological evolution of cosmic ray sources can be parametrized
by power law with the sharp cut-off at some epoch with redshift
$z_{max}$ ($m$ and $z_{max}$ are considered as parameters of a
model of the combined source). At last, the function $f(E)$ in
eq.(\ref{eq5}) describes a form of the differential energy
spectrum of the local source. For a simple power law injection
spectrum we can write
$$
f(E)=\frac{\gamma - 1}{E_0} \Big(\frac{E}{E_0}\Big)^{-\gamma}.
$$

Now supposing some values for the presented parameters we can
solve (\ref{eq1}). One more thing needed is infrared photon number
density for different redshifts, $n^{IR}(E_{\gamma},z)$, to
calculate energy losses in infrared background. As for the case of
relic photons, the dependence of $n^{relic}(E_{\gamma},z)$ on $z$
is trivial: they always have black body spectrum, with temperature
$T(z)=T(0)(1+z)$.

\section{Cosmic infrared radiation background}
Extragalactic infrared background had been formed by
infrared-luminous galaxies in the late Universe. For the
calculation of CIB it is again convenient to use the cosmological
transport equation which is analogous to that used in the previous
section. The function which must be found is the number density of
infrared photons at different cosmological epochs,
$n^{IR}(E_{\gamma},z)$.

The resulting expression for the number density of infrared
photons in extragalactic space is \cite{b1}
\begin{equation}
\label{eq6} n^{IR}(E_{\gamma},z)=\int \limits_{z}^{z_{max}} dx
\left(\frac{1+z}{1+x}\right)^3 \int \frac{dL}{L} \rho(x,L)
S^{IR}\left(E_{\gamma} \frac{1+x}{1+z},L\right) \cdot
\frac{1}{E_{\gamma}} \Big| \frac{dt}{dx} \Big|,
\end{equation}
where $L$ is the luminosity of infrared-bright galaxy emitting
photons, $S^{IR}(E_{\gamma},L)$ describes the spectrum of this
radiation (it is called the spectral energy distribution),
$\rho(z,L)$ is the number density of infrared-bright galaxies with
a given luminosity.

From direct measurements we can learn  $\rho(0,L)$ and
$S^{IR}(E_{\gamma},L)$. Then we can assume some kind of the
cosmological evolution to obtain $\rho(z,L)$. Next step is to
calculate $n^{IR}(E_{\gamma},0)$ making use of formula
(\ref{eq6}), and compare it with the measured CIB intensity. One
of the results of such estimations is the well-known fact
\cite{franceschini,hauser} that we need to assume strong evolution
of background sources in the far infrared region to explain
observed CIB intensity (say, $\nu I(\nu)\sim 30nW m^{-2}sr^{-1}$
at $\lambda \sim 100\mu m$).

In fact, we include in our calculations only far and partially
middle infrared background regions, not considering near infrared
and optic regions. For the present calculations, we used the same
parametrization of the function $\rho(z,L)$ and same input data as
described in \cite{b2}. But we should mention that the effect
under discussion (CR spectrum modification) does not depend
greatly on the type of infrared sources evolution, but much more
on the CIB intensity at $z=0$ (that is, at our time).

\section{Results and conclusions}
\begin{figure}[!t]%[h]
\begin{center}
\setcaptionmargin{5mm} %%%%
\onelinecaptionstrue  %%%%
\includegraphics*[width=0.9\textwidth]{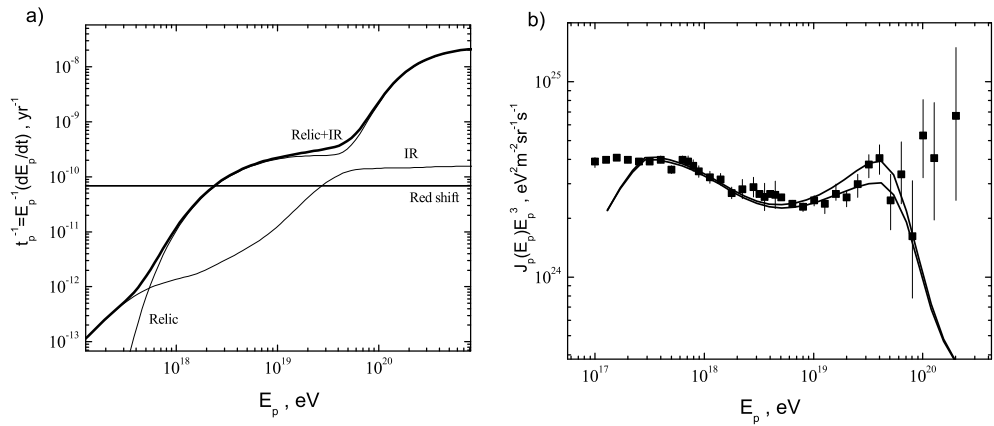}
\captionstyle{normal} %%%
\caption{{\bf a)} Proton cooling rates in relic and infrared
photon background. {\bf b)} Extragalactic CR proton spectrum
calculated with (lower curve) and without (upper curve) taking
into account energy losses in the infrared background.
\label{fig1}}
\end{center}
\end{figure}

On Fig. \ref{fig1}a we show the proton cooling rate in infrared
photon gas (together with corresponding function for the relic
photon case), both curves shown for $z=0$. The asymptotic value of
$t_p^{-1}$ at highest energies is proportional to the total number
density of photons, so for $E>10^{20} eV $ cooling on infrared
photons gives a small contribution, $\sim 0.5\%$. But one can see
from the figure that at proton energy $\sim (3\div 4)\times
10^{19} eV$, the contribution of infrared photons in total cooling
rate is noticeable.

For calculation of the CR spectrum we used the following set of
parameters determining the spectrum slope and cosmological
evolution of CR sources: $\gamma=2.5, m=3.5, z_{max}=5$. We
assumed here that extragalactic spectrum of CRs dominates
beginning from $E_0=3\times 10^{17}eV$. The theoretical CR
spectrum at $z=0$ was normalized on experimental data; from such a
normalization we obtained the value of the product $\rho_0 \eta_0$
entering eq.(\ref{eq5}) for the source function of the CR proton
kinetic equation, $\rho_0 \eta_0 \approx 1.4\times 10^{-42}
cm^{-3} s^{-1} $.

The results of two calculations of extragalactic CR proton
spectrum (with and without taking into account energy losses on
infrared photons) are shown on Fig. \ref{fig1}b. Both curves have
characteristic dip and cut-off features; one can see that
extragalactic CR spectrum near GZK cut-off gets modified when
taking into account interactions with infrared background.

In this work we used concrete model parameters for CR source
properties and the certain model for CIB, for the estimation of CR
extragalactic spectrum modification at $10^{19}-10^{20} eV$.
Certainly, new experimental data for CIB intensity as long as
improvement of our knowledge of CR sources will help to determine
more precisely the magnitude of this effect.

\end{document}